\newcommand{\bs}[1]{\mbox{\boldmath $#1$}}
\input psfig
\input epsf
\documentstyle[preprint,prb,aps]{revtex}
\begin{document}
\preprint{}
\draft
\title{Toroidal vortices in resistive magnetohydrodynamic equilibria}
\author{David Montgomery, Jason W. Bates, and Shuojun Li}
\address{Department of Physics and Astronomy \\
Dartmouth College \\
Hanover, New Hampshire $ $ 03755-3528 U.S.A.}
\date{\today}
\maketitle
\begin{abstract}
Resistive steady states in toroidal magnetohydrodynamics (MHD), where
Ohm's law must be taken into account, differ considerably from ideal
ones. Only for special (and probably unphysical) resistivity profiles
can the Lorentz force, in the static force-balance equation, be
expressed as the gradient of a scalar and thus cancel the gradient of
a scalar pressure. In general, the Lorentz force has a curl directed
so as to generate toroidal vorticity.  Here, we calculate, for a
collisional, highly viscous magnetofluid, the flows that are required
for an axisymmetric toroidal steady state, assuming uniform scalar
resistivity and viscosity. The flows originate from paired toroidal
vortices (in what might be called a ``double smoke ring''
configuration), and are thought likely to be ubiquitous in the
interior of toroidally driven magnetofluids of this type. The
existence of such vortices is conjectured to characterize
magnetofluids beyond the high-viscosity limit in which they are
readily calculable.
\end{abstract} 
\vspace{0.5in}
\narrowtext
\section{INTRODUCTION}
If static solutions of the MHD equations in toroidal geometry are
forced to obey Ohm's law as well as force balance, most of them
disappear. The reason is that the current densities resulting from 
inductively generated, steady electric fields, give the term ${\bf j} 
\bs{\times} {\bf B}$ in the equation of motion (where ${\bf j}$ is the 
electric current density and ${\bf B}$ is the magnetic field) a finite 
curl, and thus ${\bf j} \bs{\times} {\bf B}$ cannot be balanced by the 
gradient of a scalar pressure.\cite{Montgomery94} The only exceptions, 
for finite, uniform transport coefficients and incompressible MHD, involve
resistivity profiles which are not uniform on a magnetic flux surface,
and thus may be thought unrealizeable.\cite{Bates96} Earlier, it was
speculated\cite{Montgomery94} that if toroidal MHD steady states exist for
resistive magnetofluids, they will probably involve flows (vector
velocity fields).  Here, we wish to calculate and describe such
velocity fields for the case of large plasma viscosity (low viscous
Lundquist number).  We stress that the effect we are describing is a
consequence of toroidal geometry, and is not an issue in the ``straight
cylinder'' approximation. Vortices similar to the ones found here can
appear in straight-cylinder MHD computations but only above
instability thresholds.\cite{Shan93,Montgomery95} In the toroidal case, the
vortices are an integral part of the force balance in the steady state
all the time, and are not uniquely connected with instabilities.
\par
In Sec.~II, the governing MHD equations are written out, and an
approximate method of solution valid in the limit of large viscosity
(low viscous Lundquist number) is discussed. With our approach, the
velocity fields necessary to maintain axisymmetric, toroidal,
resistive steady states are determined. In Sec.~III, numerical
implementation of the procedure is described, and vorticity contours
and streamlines resulting from the calculations are presented.  We
believe the results may suggest a simultaneous occurrence of current
circuits and vortex rings in a finitely electrically conducting fluid
that goes well beyond this single tractable example. Sec.~IV is a
summary and suggestion for further investigations. The Appendix
describes, in terms of inequalities among dimensionless Reynolds-like
numbers, the approximations made and also specifies the geometry in
detail. Purely for computational convenience, we specialize to a
toroid with a rectangular cross section (see the Appendix and
Fig.~1). However, we believe our results to apply to toroids with more
general boundary shapes.

\section{A HIGH-VISCOSITY CALCULATION}

The MHD equation of motion for a uniform-density, incompressible magnetofluid 
is\cite{Shercliff65}
\begin{equation}
\frac{\partial {\bf v}}{\partial t} + {\bf v} \cdot \nabla{\bf v} = 
{\bf j} \bs{\times}{\bf B} - \nabla p + \nu \nabla^2{\bf v},
\label{eq: eqmo}
\end{equation}                                                                 
where ${\bf v}$ is the fluid velocity, $p$ is the pressure, and $\nu$ 
is the kinematic viscosity.  We work in standard ``Alfv\'{e}nic'' 
dimensionless units, so that in fact $\nu$ is the reciprocal of a 
Reynolds-like number; specifically, $\nu^{-1}$ is the viscous Lundquist 
number, given in terms of quantities expressed in $cgs$ units by
\begin{equation}
\nu = C_a L/\tilde{\nu},                                           
\end{equation}
where $C_a$ is an Alfv\'{e}n speed based on the mean poloidal magnetic
field, $L$ is a characteristic length scale which can be taken to be a
toroidal minor radius, and $\tilde{\nu}$ is the laboratory kinematic
viscosity, expressed in $cm^2/s$. We assume that
the magnetofluid is incompressible, and that the viscosity and
electrical conductivity are spatially uniform scalars.  These are all
significantly restrictive assumptions, but they will be seen to lead to
a tractable problem in otherwise uncharted territory.
\par
The most severe assumption we will make is that of small viscous
Lundquist number. For a collision-dominated plasma, with a mean free
path smaller than an ion gyroradius, $\tilde{\nu}$ is essentially the
ion mean free path times an ion thermal speed. It is also the ``ion
parallel viscosity'' for the case in which the mean free path is
greater than a gyroradius,\cite{Braginskii65,Balescu88} as it is in
the case of a tokamak fusion device.  For the current generation of
tokamaks, $\tilde{\nu}$ is an extraordinarily large number, so large
as to cast legitimate doubt on the applicability of either version of
the MHD viscous stress tensor to tokamak dynamics. Considerable
discussion and some controversy has surrounded the form and magnitude
of the appropriate viscous stress tensor to be used in tokamak MHD,
and at present, these discussions show no signs of converging.  We
must admit our own doubts about the accuracy of MHD for tokamaks with
any currently available viscous stress tensor, but for purposes of
this discussion, we make the assumption that the standard isotropic
viscous-stress tensor [used in the derivation of Eq.~(\ref{eq: eqmo})]
is applicable, but with a viscosity coefficient sufficiently large as
to make the viscous Lundquist number [$1/\nu$ in Eq.~(\ref{eq: eqmo})]
small compared to unity: {\it i.e.}, $\tilde{\nu}$ is of the order of
an ion mean free path times an ion thermal speed for the case of a
plasma. In any case, the assumption made may be satisfied in other
magnetofluids and may be considered of possible relevance to tokamaks,
at least until some more convincing approximation for the viscous
stress tensor appears.
\par
This assumption makes it possible to treat the viscous term as of the
same order as the other two terms on the right hand side of 
Eq.~(\ref{eq: eqmo}), and further makes it possible to neglect the 
inertial terms [the left hand side of Eq.~(\ref{eq: eqmo})] for 
time-independent states.  A detailed justification in terms of dimensionless 
numbers is given in the Appendix. The velocity ${\bf v}$ is then to be 
calculated in terms of ${\bf j}$ and ${\bf B}$ from the approximate 
relation                                                 
\begin{equation}
\nabla p - {\bf j} \bs{\times} {\bf B} = \nu \nabla^2 {\bf v},
\label{eq: redeqmo}
\end{equation}
where to lowest order, ${\bf j}$ and ${\bf B}$ are to be obtained 
strictly from Ohm's and Amp\`{e}re's laws and the magnetic boundary 
conditions, without reference to ${\bf v}$.  The term 
containing velocity in Eq.~(\ref{eq: redeqmo}) needs to be retained in 
order to balance the part of the left hand side that has a non-vanishing 
curl. The resistivity and viscosity are to be taken as spatially 
uniform, so that taking the divergence of Eq.~(\ref{eq: redeqmo}) would 
give, using the incompressibility assumption,
\begin{equation}
\nabla^2 p = \nabla \cdot \left( {\bf j} \bs{\times} {\bf B}\right),
\end{equation}
a Poisson equation for the pressure that determines $p$ as a functional
of ${\bf j}$ and ${\bf B}$.  The pressure can also be made to drop out of 
Eq.~(\ref{eq: redeqmo}) by taking the curl and writing the result as an 
inhomogeneous equation to be solved for the vorticity 
$\bs{\omega} = \nabla \bs{\times} {\bf v}$:
\begin{equation}
\nabla \bs{\times} \left({\bf j} \bs{\times} {\bf B}\right) = - \nu 
\nabla^2\left(\omega_\varphi \hat{{\bf e}}_\varphi\right) 
= - \nu \nabla^2 \bs{\omega}.
\label{eq: Peqvort}
\end{equation}
We shall find that in the geometry considered, the vorticity vector points 
entirely in the (toroidal) $\varphi$-direction; see the Appendix.
Once the fluid velocity and the vorticity are determined by solving
Eqs.~(\ref{eq: redeqmo}) and (\ref{eq: Peqvort}) (${\bf v}$ and 
$\bs{\omega}$ are, as advertised, ``small'' in the sense
of being first order in $1/\nu$), one can return with them to Ohm's
and Amp\`{e}re's laws and iterate again, obtaining first-order corrections
to the current and magnetic field. Then, corrections to ${\bf v}$
and $\bs{\omega}$ can be obtained by going back again to 
Eqs.~(\ref{eq: redeqmo}) and (\ref{eq: Peqvort}), and iterating.  
Our interest here, though, is in the lowest order solutions for the 
velocity and vorticity.
\par
The electric field, highly idealized for tractability, is regarded as
being generated by a time-proportional axial magnetic field confined
to a high permeability cylinder (a cylindrical iron core, say) whose
axis of symmetry is the $z$-axis, and which extends to infinity in the
positive and negative $z$ directions. This cylinder lies entirely within
the ``hole in the doughnut'' of the toroid and is perpendicular to the
mid-plane $z=0$.  It produces an electric field which lies purely in the 
azimuthal direction [$\varphi$-direction in cylindrical polar coordinates 
$(r,\varphi,z)$, which we use throughout]: ${\bf E}=\left(E_0r_0/r\right)
\hat{{\bf e}}_\varphi$, where $E_0$ is the strength of the applied electric 
field at a reference radius $r=r_0$ within the toroid.  Ohm's law, in 
the dimensionless units, is (with a resistive Lundquist number $1/\eta$)
\begin{equation}
{\bf E} + {\bf v} \bs{\times} {\bf B} = \eta {\bf j},
\label{eq: Ohmlaw}
\end{equation}
so that if we neglect ${\bf v}$ to lowest order, a purely toroidal 
current density is generated with the form 
${\bf j}=\left(E_0r_0/\eta r\right)\hat{{\bf e}}_\varphi$.  The (poloidal)
magnetic field associated with this ${\bf j}$ is determined by
Amp\`{e}re's law and by boundary conditions, which we take to be ${\bf
B} \cdot \hat{\bf n} = 0$ at the toroidal walls, where $\hat{{\bf n}}$
is the unit normal.  We assume that the toroidal boundaries are highly 
conducting and coated with a thin layer of insulator; we ignore the slits and
slots in the conducting walls that are required for the applied electric
field to penetrate the interior of the toroid: a necessary and common
if regrettable idealization. We return presently to the explicit
calculation of the poloidal magnetic field.  In addition, we may
assume the presence of a vacuum toroidal magnetic field that is
externally supported: ${\bf B}_T=B_\varphi \hat{{\bf e}}_\varphi=
\left(B_0r_0/r\right)\hat{{\bf e}}_\varphi$. The total magnetic field is 
the sum of the toroidal and poloidal magnetic fields.  The toroidal
magnetic field seems to play little role in establishing the
properties of the equilibrium, though it will have a great deal to say
about the stability of that equilibrium, a question which we do not
consider here. The magnetic field lines are, topologically speaking,
helical; but the toroidal field enters in no other context, and the
poloidal magnetic field contains all the nontrivial magnetic
information. The toroidal magnetic field would also play a much more
prominent role if a tensor electrical conductivity were allowed. A
tensor electrical conductivity, which we do not consider, would permit
lowest-order poloidal currents as well; this limitation would be
desirable to remove in future treatments. At the next order, poloidal
currents would also be implied by the cross product of ${\bf v}$ with
the toroidal magnetic field. It is to be stressed that until the
velocity field begins to feature in the calculation, the mathematics
are indistinguishable from a similar low-frequency electrodynamics
calculation in a finitely conducting ring of metal.  It is shown
in the Appendix that the condition for the negligibility of ${\bf v}$
in Ohm's law at the lowest order is again the smallness of the viscous
Lundquist number, not a totally obvious result.
\par        
Before computing the poloidal magnetic field explicitly, we return to 
Eq.~(\ref{eq: Peqvort}) and examine how the vorticity $\bs{\omega}$ 
may be determined once the left hand side is known. Since the Laplacian 
of a vector in the toroidal direction that only depends on poloidal 
coordinates is also a vector in the toroidal direction that only depends 
on poloidal coordinates, we essentially have a Poisson equation to solve 
for $\bs{\omega} = \omega_\varphi \hat{{\bf e}}_\varphi$.  One way to proceed 
is to expand $\bs{\omega}$ in a family of vector eigenfunctions of the 
Laplacian, which are related to waveguide modes:
\begin{equation}
\nabla^2\left(\omega_\varphi\hat{{\bf e}}_\varphi\right) 
+ \lambda^2 \left(\omega_\varphi\hat{{\bf e}}_\varphi\right)= 0.
\label{eq: vecLap}
\end{equation}   
All components of all fields are axisymmetric ({\it i.e.}, 
the components of all vector fields are $\varphi$-independent), and the 
solution to Eq.~(\ref{eq: vecLap}) is any one of the functions:
\begin{equation}
\omega_{jk}\hat{{\bf e}}_\varphi \equiv 
\varepsilon_{jk}\left[J_1(\gamma_{jk}r)+D_{jk}Y_1(\gamma_{jk}r) \right] 
\left( \begin{array}{c} \sin kz \\ \cos kz \end{array} \right)
\hat{{\bf e}}_\varphi,
\label{eq: eigenfunc}
\end{equation}
where $\varepsilon_{jk}$, $\gamma_{jk}$, $D_{jk}$ and $k$ are
undetermined constants, with $\gamma_{jk}$, $k$ and $\lambda_{jk}$
related by the condition $\lambda^2_{jk}= \gamma^2_{jk}+k^2$. Here,
$J_1$ and $Y_1$ are Bessel and Weber functions, respectively. The
vorticity may be written in terms of a velocity stream function
$\psi(r,z)$ in the following way: $\omega_\varphi \hat{{\bf
e}}_\varphi = - \nabla^2\left(\psi \hat{{\bf e}}_\varphi/r \right)$. 
The function $\psi/r$ may also be expanded 
in terms of the eigenfunctions obtained from Eqs.~(\ref{eq: vecLap})
and (\ref{eq: eigenfunc}).  The boundary conditions chosen are that
$\psi$ and $\omega_\varphi$, and thus the $\omega_{jk}$, vanish on the
boundary.  We now choose a rectangular cross section for the toroid;
the combined boundary conditions ${\bf v}\cdot \hat{{\bf n}}=0$ and
$\omega_\varphi=0$ make the velocity field satisfy stress-free,
impenetrable boundary conditions. That is, ${\bf v}\cdot \hat{{\bf
n}}$ and the tangential viscous stress are zero on all four
faces. These are the only boundary-shape and internally consistent
viscous boundary conditions (perfectly smooth and impenetrable walls)
we have found for which the solution is obtainable by elementary
means. However, we believe the qualitative conclusions to be reached
will apply to essentially an arbitrary toroidal boundary shape.
\par
We seek a solution to Eq.~(\ref{eq: Peqvort}) by expanding
$\omega_\varphi$ in terms of the eigenfunctions, $\omega_{jk}$, having
determined numerically the allowed values of $\varepsilon_{jk}$, 
$\gamma_{jk}$, $D_{jk}$ and $k$.  The value of $\varepsilon_{jk}$ is 
chosen so that the two-dimensional integral of the square of the
eigenfunctions is unity. We assume that what we have obtained, then, 
is a complete orthonormal set.  We write
\begin{equation}
\omega_\varphi=\sum_{j,k}
\Omega_{jk}\varepsilon_{jk}\left[J_1(\gamma_{jk}r)
+D_{jk}Y_1(\gamma_{jk}r)\right] \left( 
\begin{array}{c} \sin kz \\ \cos kz \end{array} \right)
\equiv \sum_{j,k}\,\Omega_{jk}\,\omega_{jk},
\label{eq: psiphi}
\end{equation}
with unknown expansion coefficients $\Omega_{jk}$. Assuming term-by-term
differentiability of the expression in Eq.~(\ref{eq: psiphi}), the 
stream function is given by
\begin{equation}
\psi = r\,\sum_{j,k}\lambda^{-2}_{jk}\,\Omega_{jk}\,\omega_{jk}.
\end{equation}                                                                 
Using the orthonormal eigenfunctions $\omega_{jk}$, Eq. (5) may be expressed 
as
\begin{equation}
\nabla \bs{\times} \left({\bf j} \bs{\times} {\bf B}\right) = 
\nu \,\sum_{j,k}\lambda^2_{jk} \, \Omega_{jk} \, \omega_{jk}\,
\hat{{\bf e}}_\varphi.
\end{equation}                                                               
If we choose, we may obtain ${\bf B}$ from ${\bf j}$ by writing it in terms 
of a vector potential ${\bf B}= \nabla \bs{\times {\bf A}}$, where 
${\bf A}=A_\varphi \hat{{\bf e}}_\varphi$ is to be expanded using the same 
eigenfunctions as we used to expand the vorticity and the stream function. 
Since the curl of $\nabla \bs{\times} {\bf A} = {\bf B}$ yields 
(if $\nabla \cdot {\bf A}=0$)
\begin{equation}
\nabla^2\left(A_\varphi \hat{{\bf e}}_\varphi \right) = -j_\varphi 
\hat{{\bf e}}_\varphi = -\frac{E_0r_0}{\eta \, r}\hat{{\bf e}}_\varphi,
\label{eq: anPo}
\end{equation}                                             
it is clear that ${\bf A}$ and ${\bf B}$ can be determined by
the previous procedure: solving Eq.~(\ref{eq: anPo}) by expanding in
the same eigenfunctions (${\bf A}= 0$ on the wall) and
differentiating.  In the next section, we adopt an alternative and
slightly simpler procedure to determine ${\bf B}$.  It is already
clear, however, what the form is that the vorticity field will have to
take, simply by looking at the source term in the Poisson equation
for vorticity and noting the fact that the radial component of ${\bf
B}$ will be positive above the mid-plane and negative below.  The
vorticity distribution will look dipolar in the ($r,z$)-plane, and the
three-dimensional vortices will be vortex rings that sit one on top
of the other, above and below the mid-plane: a ``double smoke ring''
configuration.

\section{EXPLICIT NUMERICAL SOLUTIONS}

In this section, we determine the poloidal vector fields ${\bf B}_p$
and ${\bf v}$ for this MHD system. These fields are associated with a
toroidal current density that is proportional to $1/r$, where $r$ is
the distance from the toroidal axis. For illustrative purposes, we
consider a toroid with a rectangular cross section (see Fig.~1).  We
assume the boundaries to lie at the planes $z= \pm L$ and at the radii
$r=r_-$ and $r=r_+$, where $r_- \leq r \leq r_+$. In our calculations, all
length scales are normalized to the major radius $r_0 \equiv (r_- +
r_+)/2$.  Since we do not assume large aspect ratio, there is no 
significant difference in the lengths of the major and minor radii. 
Because the walls of the toroid are idealized as perfect
conductors, the appropriate magnetic boundary condition for this problem is
that the normal component of the magnetic field vanish there. To find
the magnetic field we solve for the magnetic flux function $\chi$ directly. 
In terms of $\chi$, both the vector potential and the poloidal magnetic 
field can be easily derived. Once ${\bf B}_p$ is known, the source term $\nabla
\bs{\times} ({\bf j} \bs{\times} {\bf B})$ on the left side of Eq.~(5)
can be computed, and the resulting Poisson equation solved for the
vorticity. With the vorticity, the velocity stream function, $\psi$,
and finally the velocity ${\bf v}$ can also be found.
\par
With axisymmetry, the poloidal component of the magnetic field ${\bf B}_p$ 
may be represented in terms of a magnetic flux function $\chi(r,z)$ according 
to
\begin{equation}
{\bf B}_p = \nabla \chi \;\; \bs{\times} \;\; \nabla \varphi.
\label{eq: Bpoloidal}
\end{equation} 
Substituting this into Amp\`{e}re's law, $\nabla \bs{\times} {\bf B}
= {\bf j}$, with ${\bf j} = \left(E_0r_0/\eta r\right) 
\hat{{\bf e}}_\varphi$ yields
\begin{equation}
\Delta^* \chi = r \frac{\partial }{\partial r} \frac{1}{r} 
\frac{\partial \chi}{\partial r} + \frac{\partial^2 \chi}{\partial z^2} = - 
\frac{E_0r_0}{\eta}.
\label{eq: delstar}
\end{equation}
Note that the magnetic vector potential ${\bf A} = A_\varphi 
\hat{{\bf e}}_\varphi$ is obtained from the magnetic flux function by 
dividing by $r$: $A_\varphi = \chi/r$. We seek $\chi$ and ${\bf B}_p$ 
in the rectangular domain $r_- \leq r \leq r_+$ and $-L \leq z \leq L$ 
subject to the boundary condition ${\bf B} \cdot \hat{{\bf n}} = 0$, where 
$\hat{{\bf n}}$ is the unit normal to the wall of the toroid. This boundary 
condition implies
\begin{mathletters}
\label{eq: homoall}
\begin{equation}
\frac{\partial \chi}{\partial r} = 0, \hskip 0.2in {\rm at} 
\hskip 0.2in z = \pm L\,\;\;\;\; \label{eq: bc1}
\end{equation}
\begin{equation}
\frac{\partial \chi}{\partial z} = 0, \hskip 0.2in {\rm at}
\hskip 0.2in r = r_+, r_-\,. \label{eq: bc2}
\end{equation}
\end{mathletters}
A particular solution of Eq.~(\ref{eq: delstar}) that vanishes at $r=r_-$ 
and $r=r_+$ is
\begin{equation}
\chi_p = - \frac{E_0r_0}{2\eta}\left\{ r^2\ln\frac{r}{r_-}-
\frac{r^2-r^2_-}{r^2_+-r^2_-}\;r^2_+\;\ln\frac{r_+}{r_-} \right\}.
\end{equation}
The solution of the homogeneous equation $\Delta^*\chi_h = 0$ that is 
symmetric about the mid-plane of the toroid is
\begin{equation}
\chi_h = \sum_\kappa C_\kappa \epsilon_\kappa r\left[J_1(\kappa r)+
D_\kappa Y_1(\kappa r)\right]\cosh \kappa z,
\end{equation}
where $C_\kappa$, $\epsilon_\kappa$, $D_\kappa$ and $\kappa$ are 
arbitrary constants. The general solution to Eq.~(\ref{eq: delstar}) 
is $\chi = \chi_p+\chi_h$.
\par
Equation (\ref{eq: bc2}) is satisfied by requiring that 
\begin{eqnarray}
J_1(\kappa r_-) + D_\kappa Y_1(\kappa r_-) &=& 0\, , \label{eq: conrm} \\
J_1(\kappa r_+) + D_\kappa Y_1(\kappa r_+) &=& 0\, , \label{eq: conrp}
\end{eqnarray}	
Equations (\ref{eq: conrm}) and (\ref{eq: conrp}) can only be 
solved consistently if the determinant 
\begin{equation}
{\cal D} \equiv J_1(\kappa r_-)Y_1(\kappa r_+) - J_1(\kappa r_+)Y_1(\kappa r_-)
\end{equation}	
vanishes. For an infinite sequence of $\kappa$-values, with each $\kappa$
corresponding to a particular zero of ${\cal D}$ for given values of
$r_-$ and $r_+$, general Sturm-Liouville theory tells us that the
functions
\begin{equation}
\phi_{0 \kappa} \equiv \epsilon_{\kappa} \left[J_0(\kappa r) + D_\kappa\,
Y_0(\kappa r)\right],
\end{equation}	
form a complete orthonormal set on the interval $r_- \leq r \leq r_+$.  The 
$\epsilon_\kappa$ are real constants chosen to normalize the 
$\phi_{0 \kappa}$:
\begin{equation}
\int_{r_-}^{r_+} \phi_{0 \kappa} \phi_{0 \kappa'} r\,dr = \delta_{\kappa,\,
\kappa'}\,.
\end{equation}								
\par
The $z$-boundary condition can be satisfied by requiring that
\begin{equation}
\frac{E_0r_0}{2\eta}  \left\{1 + 2 \ln\frac{r}{r_-}-\frac{2\,r_+^2}
{r^2_+ - r^2_-}\ln\frac{r_+}{r_-} \right\} = 
\sum_\kappa \kappa C_\kappa \epsilon_\kappa \left[J_0(\kappa r) + D_\kappa 
\,Y_0(\kappa r)\right]\cosh \kappa L.
\end{equation}
Multiplying both sides of this equation by $r\, \phi_{0 \kappa'}(r)$ and 
integrating from $r_-$ to $r_+$ determines the coefficients $C_\kappa$. 
We find
\begin{equation}
C_\kappa = \frac{E_0r_0}{2\eta} \int_{r_-}^{r_+} \phi_{0 \kappa}(r)\left\{1
+2\ln\frac{r}{r_-}-\frac{2r^2_+\ln(r_+/r_-)}{r^2_+-r^2_-}\right\}rdr/
(\kappa \cosh \kappa L).
\end{equation}
Table~I shows the first ten values of $r_0\,\kappa$, 
$D_\kappa$, $r_0 \, \epsilon_{\kappa}$ and $2\,\eta\,C_\kappa/(E_0\,r_0^3)$ 
for $r_-/r_0=0.6$, $r_+/r_0=1.4$, and $L/r_0=0.3$. With these coefficients 
and the discrete set of $\kappa$-values, a magnetic flux function $\chi$ 
that satisfies both boundary conditions in Eq.~(\ref{eq: homoall}) 
can be constructed. A plot of the contours of $\chi(r,z)$ appears in 
Fig.~2. These contours are the projections of the surfaces on which the 
magnetic field lines lie. Using Eq.~(\ref{eq: Bpoloidal}), the components 
of the poloidal magnetic field can be easily computed. We find
\begin{eqnarray}
B_r(r,z) &=& \sum_\kappa C_\kappa \epsilon_\kappa \kappa \left[J_1(\kappa r)
+D_\kappa Y_1(\kappa r)\right] \sinh \kappa z, \label{eq: Br} \\
B_z(r,z) &=& -\frac{E_0r_0}{2\eta}  \left\{1 + 2 \ln\frac{r}{r_-}
-\frac{2r_+^2}{r^2_+ - r^2_-}\ln\frac{r_+}{r_-} \right\} \nonumber \\
\mbox{} & & + \sum_\kappa  C_\kappa \epsilon_\kappa \kappa \left[J_0(\kappa r)
+D_\kappa Y_0(\kappa r)\right] \cosh \kappa z. \label{eq: Bz}
\end{eqnarray}
\par
The next step in our derivation is to determine the vorticity, 
$\omega_\varphi$. For a current density $j_\varphi = E_0r_0/\eta r$, the 
left side of Eq.~(11) reduces to $-\left(2E_0r_0B_r/\eta r^2\right)
\hat{{\bf e}}_\varphi$. Thus, the equation to solve is
\begin{equation}
\sum_{n,\ell}\lambda^2_{n\ell}\,\Omega_{n\ell}\,\omega_{n\ell} = - 
\frac{2E_0r_0}{\eta\nu}\frac{B_r}{r^2},
\label{eq: perhaps} 
\end{equation}
where $B_r$ is given by Eq.~(\ref{eq: Br}), and the expansion coefficients
$\Omega_{n\ell}$ are as of yet undetermined. The eigenfunctions 
$\omega_{n\ell}$ that have odd parity in $z$ and vanish on the boundary 
of the toroid are given by
\begin{equation}
\omega_{n\ell}(r,z)=\frac{1}{\sqrt{L}}\;\phi_{1n}(r)\sin 
\frac{\ell\pi z}{L}, \hspace{0.5in} \ell = 1,2,3, \ldots ,
\end{equation}
where $1/\sqrt{L}$ is a normalization factor. Possible 
$\cos[(2\ell+1)\pi z/2L]$ terms may be omitted from symmetry 
considerations. The functions $\phi_{1n}(r)$ are defined by
\begin{equation}
\phi_{1n}(r) \equiv \varepsilon_{n}\left[J_1(\alpha_n r)+D_nY_1(\alpha_n r)
\right],
\end{equation}
where the parameters $\varepsilon_{n}$ are real constants chosen to normalize 
$\phi_{1n}$:
\begin{equation}
\int_{r_-}^{r_+} \phi_{1n} \phi_{1n'} \,r\,dr = \delta_{n,\,n'}\,.
\end{equation}		
The ``$n$'' in $\alpha_n$ designates the $n$-th zero of the function 
$\cal{D}$, and $\alpha_n^2=\lambda^2_{n\ell}-\ell^2\pi^2/L^2$. Note 
that the $\alpha_n$-values correspond to the $\kappa$-values, the first 
ten of which appear in Table I for $r_-/r_0 = 0.6$, $r_+/r_0=1.4$ and 
$L/r_0=0.3$. Multiplying Eq.~(\ref{eq: perhaps}) by 
$\omega_{n',\ell'}\,rdrdz$ and 
integrating over the range $r_- \leq r \leq r_+$ and $-L \leq z \leq L$ 
determines the expansion coefficients $\Omega_{n\ell}$. The result is
\begin{eqnarray}
\Omega_{n\ell} &=& \frac{4\pi E_0 r_0\,\ell(-1)^{\ell}}{\eta\nu L^{3/2}
\left(\alpha^2_n+\ell^2\pi^2/L^2\right)}\sum_m  
\frac{C_m\,\alpha_m\,\sinh \alpha_m L}{\alpha^2_m+\ell^2\pi^2/L^2} \nonumber \\
\mbox{} & & \hspace{2in} \times \int_{r_-}^{r_+}\phi_{1m}(r)
\phi_{1n}(r)\frac{dr}{r}.
\end{eqnarray}
\par
A contour plot of $\omega_\varphi$ using these coefficients appears in
Fig.~3 for $r_-/r_0=0.6$, $r_+/r_0=1.4$ and $L/r_0=0.3$. Positive contours
are denoted by a solid line and negative contours by a dashed line. The 
vorticity vanishes at the toroidal walls, which is
equivalent to stress-free boundary conditions in this problem. The
convergence of the vorticity series is rather fast, owing to the
presence of $\alpha^2_n+\ell^2\pi^2/L^2$ terms in the denominator of
$\Omega_{n\ell}$.  Typically, it was only necessary to keep a dozen or
so terms to achieve a high degree of accuracy. Contours of the velocity 
stream function $\psi$ are shown in Fig.~4 for the same parameters and with
the same convention. With $\psi$, the velocity can be computed from 
${\bf v} = \nabla \psi \bs{\times} \nabla \varphi$. It is easily verified 
that the normal component of the velocity will vanish at the boundaries. 
In Figs.~3 and 4, note the appearance of paired-vortex structures that 
resemble a ``double smoke ring'' configuration.
\par
Since $B_r(r,z)$ in nonzero at $z= \pm L$, a Gibbs phenomenon is to be
expected in the series for $\nabla^2 \omega_\varphi \hat{{\bf
e}}_\varphi$ near $z= \pm L$. This is a consequence of representing
the $z$-dependence of the right side of Eq.~(\ref{eq: perhaps}) in
terms of sine functions, all of which vanish at $z = \pm L$. Since
both sides of Eq.~(\ref{eq: perhaps}) are identically zero at $r=r_-$
and $r=r_+$, a Gibbs phenomenon will not occur near the boundaries in
$r$.

\section{DISCUSSION AND CONCLUSIONS}
        
It is regrettable that nearly all experiments performed during
the last several years on MHD in toroidal geometry have been carried
out in tokamaks intended to confine a thermonuclear plasma.  Not only
are diagnostics for such internal variables as fluid velocity,
vorticity, and electric current density very limited due to the high
temperatures, the detailed applicability of MHD itself is in doubt
because of previously-mentioned uncertainties in the appropriate 
viscous stress tensor to be used in theoretical models. Here, we have 
attempted to isolate an interesting MHD effect, without taking a
position on whether it should or should not be an important feature of
tokamak operation.  Tokamaks are often thought to have both
overall poloidal and toroidal rotation, which have been attributed to
various consequences of local charge non-neutrality.  The combinations
of all three kinds of flows, if they were present, might be quite
difficult to untangle.
\par
The ``double smoke ring'' configuration identified in this paper is a 
feature associated with electrically-driven toroidal magnetofluids that we 
believe is quite robust; it does not require local charge non-neutrality, 
and may even appear in liquid metal experiments (for example), even though 
the high-viscosity calculations we have done imply inequalities that may not 
be easily satisfied in liquid metals.  Entirely as a consequence of the 
toroidal geometry, a purely toroidal electric current generates a magnetic 
field for which a part of the ${\bf j} \bs{\times} {\bf B}$ Lorentz force 
produces a local toroidally-directed torque on the magnetofluid. (This torque 
disappears in the ``straight-cylinder'' limit.) This gives rise to 
opposing pairs of vortex rings with vorticity aligned parallel and
anti-parallel to the current density. We believe that these structures will 
exist under a variety of boundary conditions (non-conducting walls,
for example, with no-slip boundary conditions) and will not require
low Hartmann numbers or viscous Lundquist numbers, though the flows may be 
more elaborate (involving, say, poloidal currents or toroidal velocities 
as well) when the inequalities we have invoked are not satisfied.
\par        
The likely presence of MHD flows in toroidal geometry was probably 
first reported in an unpublished paper by Pfirsch and Schl\"{u}ter
\cite{Pfirsch62} over thirty years ago. Their approach was quite different 
from ours, involving for example an inverse aspect ratio expansion. In
addition, they ignored the velocity field in the equation of motion
(but not in Ohm's law). On the basis of their model, Pfirsch and
Schl\"{u}ter concluded that there would be a necessary mass flux
outward from the toroid that required ``sources'' of mass inside the
toroid. Here, we have explicitly exhibited a large class of
solutions with no normal component of velocity at the walls, which
contradicts the findings of Pfirsch and Schl\"{u}ter.  Nevertheless,
credit for the observation that flows are to be expected in the steady
state must go to them. {\it Ideal} toroidal vortices have been considered 
by Marnachev.\cite{Marnachev87}
\par        
The flow pattern that we have been computed in this paper, with streamlines 
that cross the toroid near the mid-plane, is not one that is very propitious 
for plasma confinement with high temperatures in the center of the toroid,
and lower temperatures near the wall. Nevertheless, the ability to
``stir'' the interior of a toroidal magnetofluid with externally-maintained 
electric fields might have other applications of some interest, such as in the
cooling of alloys.\cite{Blinov89} More generally, the likely separation of 
driven MHD states into those involving velocity fields and those which are 
static seems artificial to us.  In this particular example, we have found
velocity fields that do not arise because of any instability or
malfunction, but are an inherent part of the equilibrium itself, even
though no external pressure gradients are applied.

\acknowledgements

This work was supported in part by the U.S.~Department of Energy 
under grant DE-FGO2-85ER53194.

\appendix
\section*{Reynolds-like Numbers and Geometry}
In this Appendix, we consider in detail the inequalities that justify the
neglect to lowest-order of ${\bf v} \cdot \nabla {\bf v}$ and 
${\bf v} \bs{\times} {\bf B}$ in Eqs.~(\ref{eq: eqmo}) and 
(\ref{eq: Ohmlaw}), respectively.  We also specify the assumed toroidal 
geometry of the fields ${\bf B}$, ${\bf j}$, ${\bf v}$, and $\bs{\omega}$.
\par
Proceeding from the dimensional ($cgs$ units) version of Eq.~(\ref{eq:
eqmo}), the condition for neglecting the ${\bf v} \cdot \nabla {\bf v}$ 
term relative to the viscous term is low viscous Reynolds 
number. This Reynolds number is defined as
$vL/\tilde{\nu}$, where $v$ is a typical fluid velocity, $L$ is a
typical length scale, and $\tilde{\nu}$ is a kinematic viscosity.
We take the minor toroidal radius as the typical length scale for this 
problem. Using the dimensional version of Eq.~(\ref{eq: Peqvort}), we 
may estimate $v$ as $C_a M$, where $C_a$ is an Alfv\'{e}n speed based 
on a typical poloidal magnetic-field strength $B$, and where $M$ is the 
viscous Lundquist number, $C_a L/ \tilde{\nu}$.  Inserting this in the
viscous Reynolds number requirement, we see that the justification of
Eq.~(\ref{eq: Peqvort}) follows from the smallness of the square of
$M$ compared to unity.
\par
Neglect of the velocity term in Eq.~(\ref{eq: Ohmlaw}) is justified by
requiring the velocity $v$ to be small compared to $\eta/L$. The
parameter $\eta$ is the magnetic diffusivity, defined by $\eta =
c^2/4\pi \sigma$, where $\sigma$ is the electrical conductivity. Using
the previous estimate for $v$, we see that this inequality is
equivalent to $S M << 1$, where $S$ is the resistive Lundquist number,
$C_a L/\eta$. Since $SM \equiv H^2$, where $H$ is the Hartmann number,
an equivalent statement of the second inequality is that $H^2$ is much
less than one.  Thus, the smallness of the squares of $M$ and $H$
compared to unity is enough to justify the approximations made.
\par
Throughout this paper, we work exclusively in cylindrical polar coordinates 
$(r,\varphi,z)$.  The axis of symmetry is the $z$-axis, and the mid-plane of 
the toroid is $z = 0$.  The $\varphi$-direction is the ``toroidal'' direction, 
and the $r$ and $z$ directions are called the ``poloidal'' directions. 
(See Fig.~1.) For the numerical solutions presented in Sec.~III, 
we consider a toroid with a rectangular cross section in the $(r,z)$-plane.  
The boundaries are idealized as perfectly smooth, perfectly conducting 
rigid walls, with zero normal velocity and zero tangential viscous stress.  
We imagine that the inside surfaces of the boundaries are coated with an 
infinitesimally thin layer of insulator (other idealizations 
are possible, such as a purely non-conducting boundary). In most 
experimental toroidal devices, gaps are present in the 
conducting boundary to permit the penetration of a toroidal electric field.  
The asymmetric effects introduced by these slits and slots, however, are not 
included in our model. An externally-supported toroidal (vacuum) magnetic 
field is possible, but will not feature in the analysis until a stability 
calculation or dynamical simulation is performed.  The nontrivial (poloidal) 
magnetic field, ${\bf B}_p$, therefore has only $r$ and $z$ components.  The 
current density ${\bf j}$ has only a $\varphi$-component, the vorticity 
$\bs{\omega}$ has only a $\varphi$-component, and the fluid velocity ${\bf v}$ 
has only $r$ and $z$ components. It is shown in Sec.~III that the mechanical 
flow is a (double) ``vortex ring'' configuration, and the electric current 
is a (single) ``current loop.'' The scalar pressure $p$ will depend only 
upon $r$ and $z$. The simple form of the fields in this model is a 
consequence of two assumptions: (i) axisymmetry; and (ii) isotropic scalar 
viscosity and conductivity. The introduction of a tensor conductivity or 
viscosity moves the problem out of the realm of present tractability.
\par
Note that the part of the ${\bf j} \bs{\times} {\bf B}$ term in 
Eq.~(\ref{eq: redeqmo}) that contains a curl is of first order in the 
inverse aspect ratio ({\it i.e.}, the ratio of minor and major radii of 
the toroidal system). The results of this paper, however, are not predicated
on the assumption that the inverse aspect ratio is small compared to one.

\pagestyle{empty}
\begin{figure}
\hbox{
\epsfbox{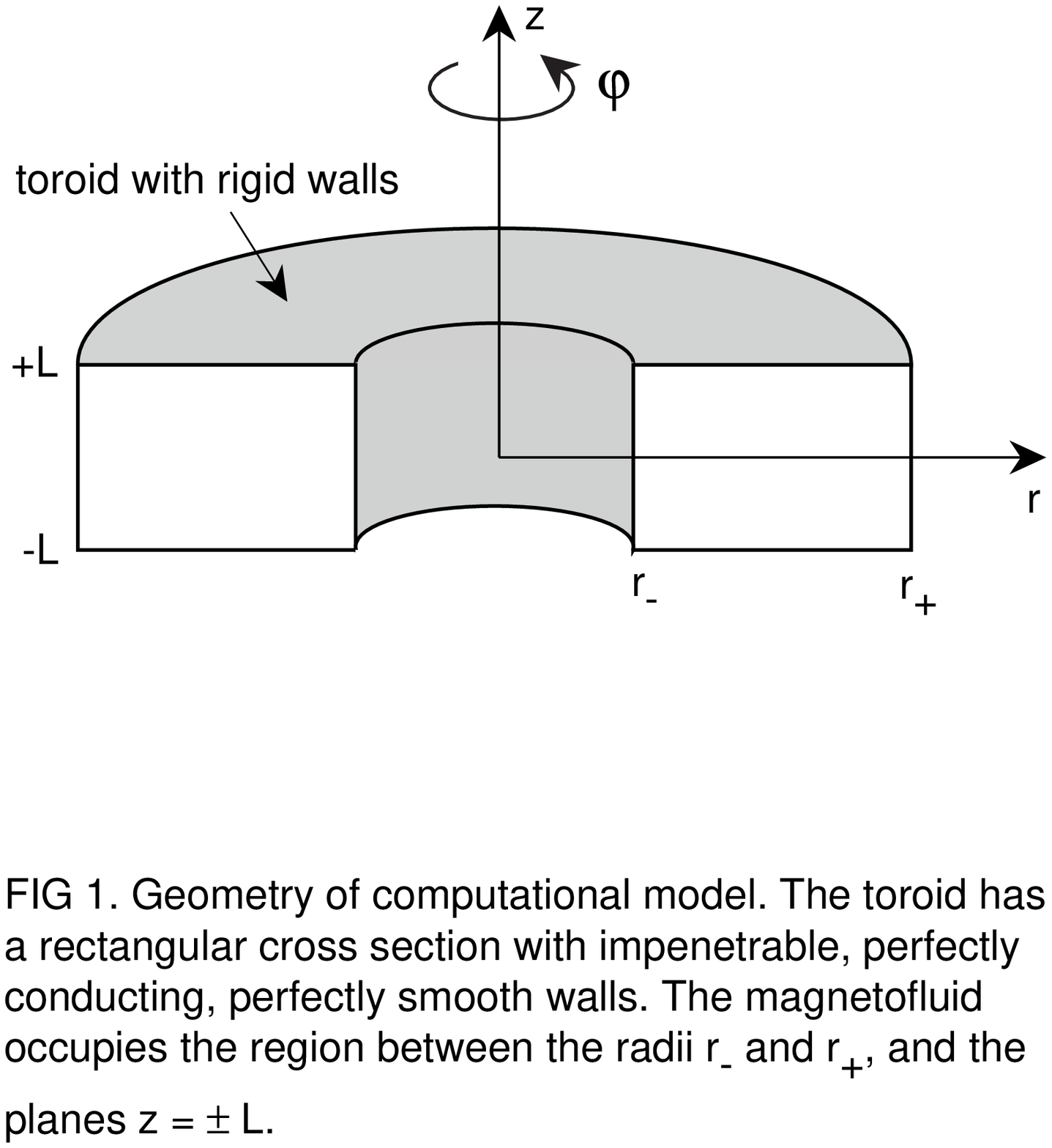}}
{\label{}}
\end{figure}

\begin{figure} 
\hbox{
\epsfbox{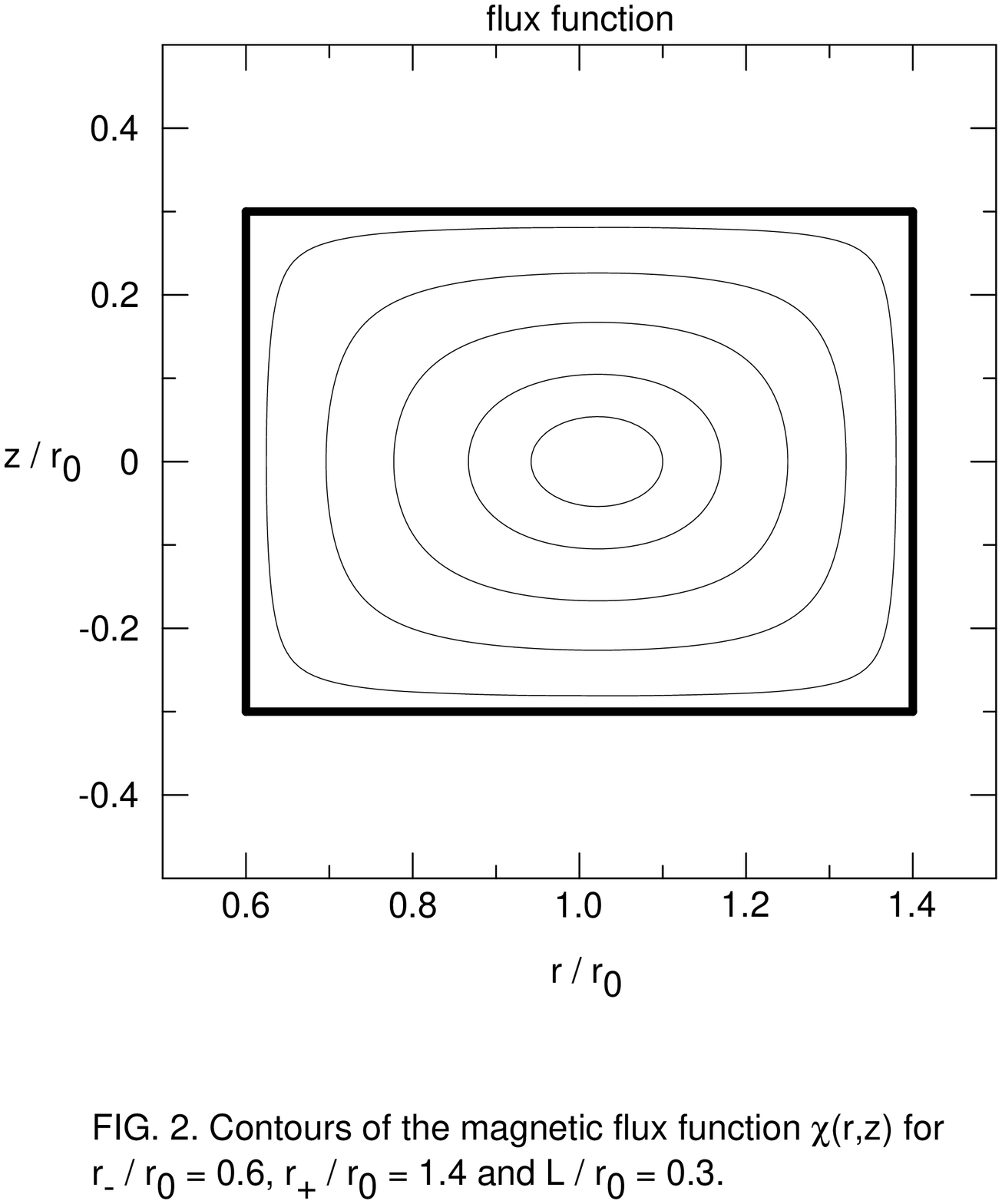}}
{\label{}}
\end{figure}

\begin{figure} 
\hbox{
\epsfbox{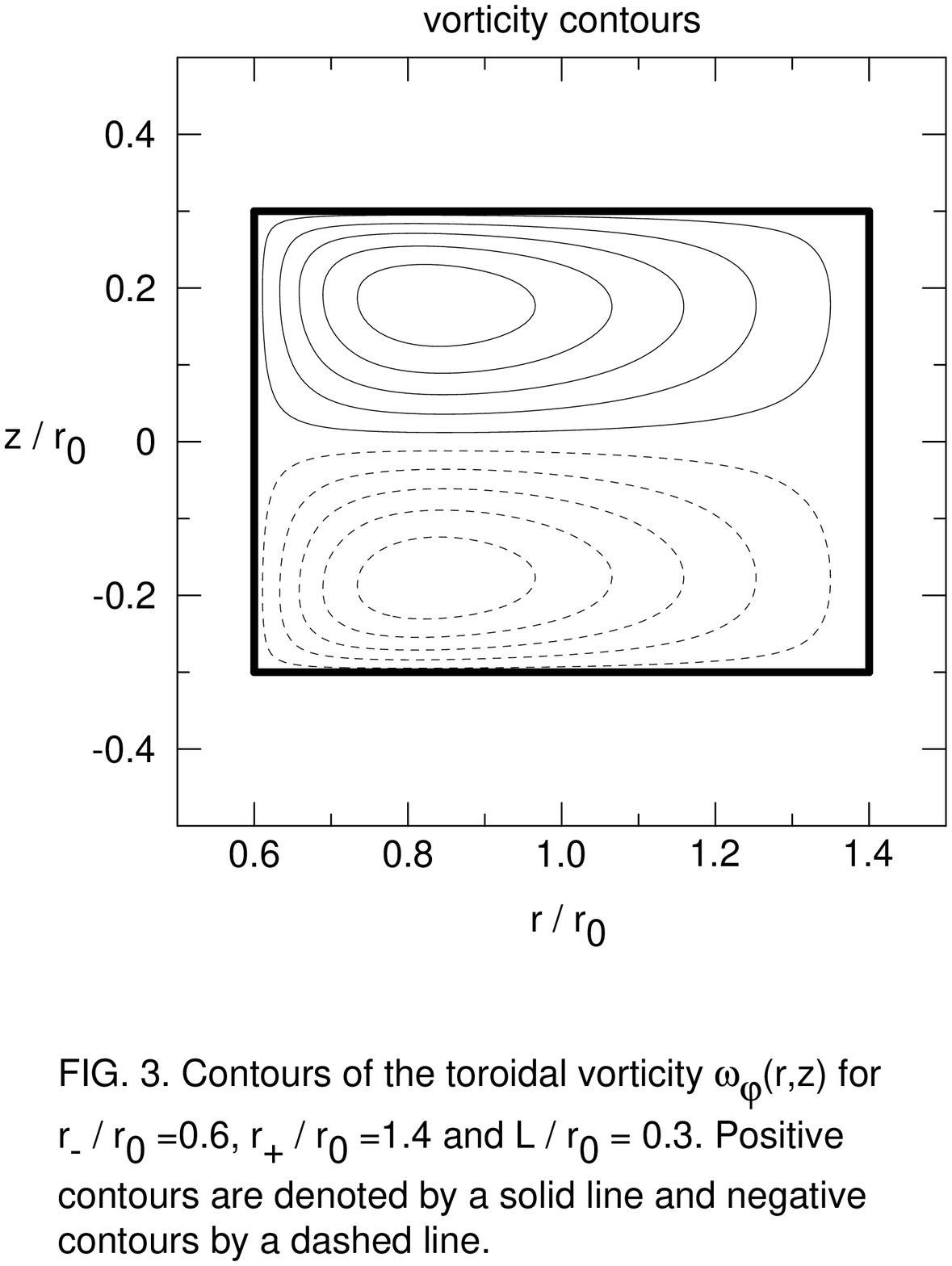}}
{\label{}}
\end{figure}

\begin{figure} 
\hbox{
\epsfbox{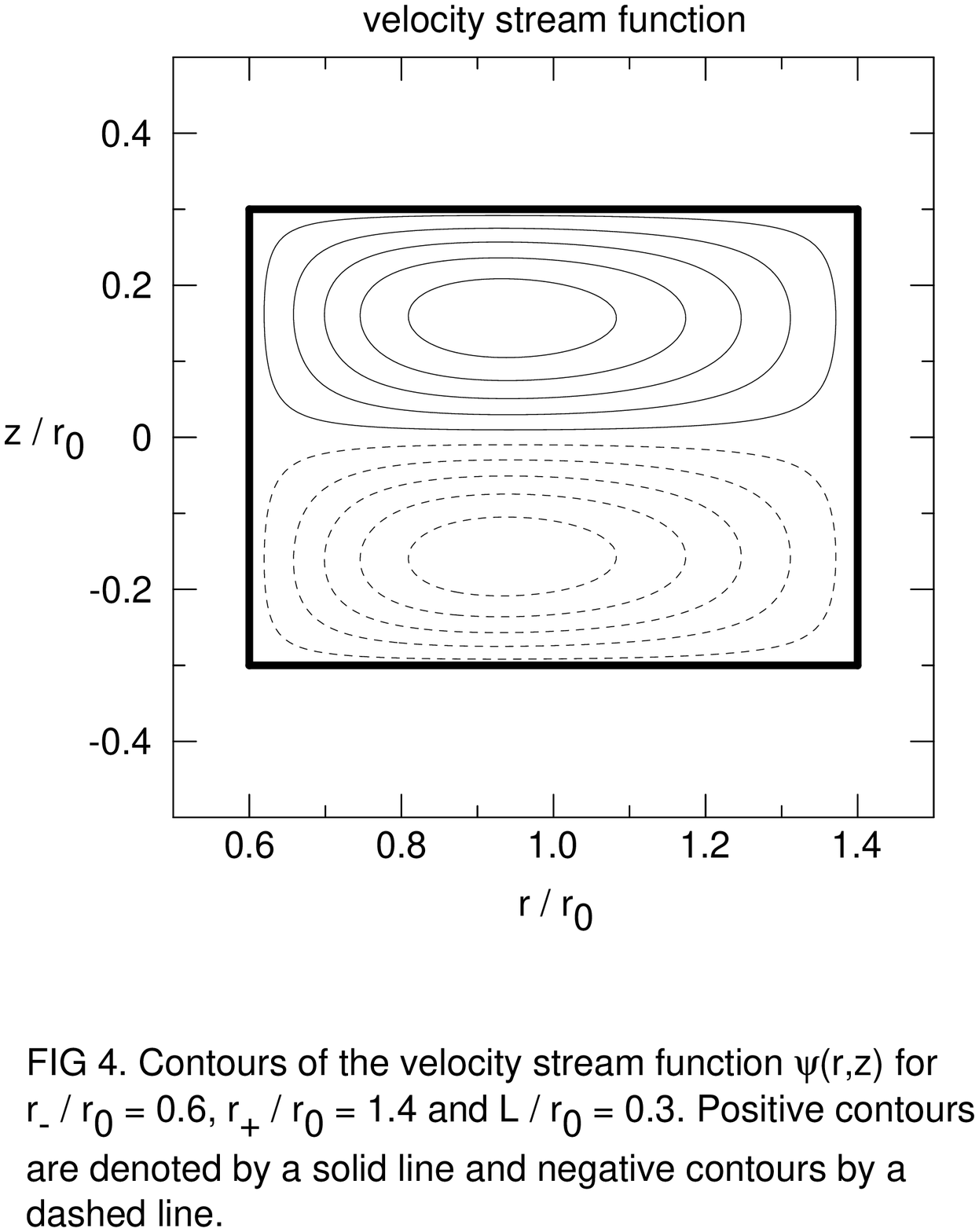}}
{\label{}}
\end{figure}

\end{document}